\documentstyle[preprint, aps]{revtex}
\begin{document}

\draft

\title{ Magnetic relaxation and critical current density of $MgB_2$ thin films}

\author{
H. H. Wen\cite{response}, S. L. Li, Z. W. Zhao, H. Jin, and Y. M. Ni  
}
\address{
National Laboratory for Superconductivity,
Institute of Physics and Center for Condensed Matter Physics,
Chinese Academy of Sciences, P.O. Box 603, Beijing 100080, China\\
}

\author{
W. N. Kang, H. J. Kim, E. M. Choi, and S. I. Lee 
}
\address{
National Creative Research Initiative Center for Superconductivity and Department of Physics, Pohang University of Science and Technology, Pohang 790-784, Republic of Korea
}

\maketitle

\begin{abstract}
Magnetic relaxation and critical current density have been measured on a $MgB_2$ thin film in a wide region of temperature with the magnetic field up to 8 T. The irreversibility line has also been determined. It is found that the relaxation rate has a very weak temperature dependence below $1/2 T_c$ showing a clear residual relaxation rate at zero K,  which cannot be easily explained as due to thermally activated flux creep. Furthermore the relaxation rate has a strong field dependence. The flux dynamics of thin films are very similar to that of high pressure synthesized bulks although the relaxation rate in thin film is systematically higher than that of a bulk sample. All the results here together with those from the bulk samples suggest that the flux dynamics may be dominated by the quantum effects, such as quantum fluctuation and tunneling.
\end{abstract}



\section{Introduction}
The recently discovered new superconductor $MgB_2$ becomes very attractive due to its potential applications\cite{akimitsu,bugoslavsky,finnemore,larbalestier}. One important issue concerns however in which region on the field-temperature ( H-T ) phase diagram it can carry large critical current density ( $j_c$ ) and thus can be used in the future for industry. This $j_c$ is controlled by the mobility of the magnetic vortices, and vanishes above the melting line between the vortex solid and liquid. This melting can be induced by strong fluctuation of the vortex position by either thermal effect or quantum effect. At T = 0 K only the quantum fluctuation is left. A finite linear resistivity $\rho_{lin}=(E/j)_{j\rightarrow 0}$ will appear above this melting point showing the reversible flux motion. Most of the published results on the flux dynamics of $MgB_2$ system were obtained from bulk samples\cite{thompson,wencpl,wenphysicac,qinmj,dhalle}. In this paper we present the investigation on the flux dynamics on an $MgB_2$ thin film by dynamical magnetic relaxation method.

\section{Experimental}
The thin films of $MgB_2$ were fabricated on (1102) $Al_2O_3$ substrates by using pulsed laser deposition technique\cite{kang}. The amorphous B thin film was first deposited and then it was sintered at a high temperature in Mg vapor. The films are typically 400 nm thick with predominant c-axis orientation ( c-axis is perpendicular to the film surface ). A rectangular sample of size $2.1mm \times 4.9 mm $ was chosen for the magnetic sweeping measurement. The magnetic measurements were carried out by a Quantum Design superconducting quantum interference device ( SQUID, MPMS 5.5 T ) and a vibrating sample magnetometer ( VSM 8T, Oxford 3001 ) at temperatures ranging from 2 K to $T_c$ and external field up to 8 T. For the magnetic sweeping measurement the M(H) curve was measured with different field sweeping rates ( 0.005 T/s to 0.02 T/s ) and integral time of 60 ms to 960 ms. The pressure of He gas in the sample chamber for thermal exchange was kept at 0.4 bar during the measurement. The superconducting transition is sharp with the transition temperature $T_c$ of about 38 K as observed from the temperature dependence of magnetization.
  
\section{Results}
In Fig.1 we show the magnetization hysteresis loops ( MHL ) measured at temperatures ranging from 2 K to 37 K. The symmetric MHLs observed at temperatures up to 37 K indicate the dominance of the bulk current instead of the surface shielding current. The MHLs measured at low temperatures, such as 2 K to about 10 K show dense and small flux jumps in the low field region leading to a smearing of the superconducting critical current density. It is this effect that makes the width of the magnetization $\Delta M $ below 10 K be anomalously smaller than that at 14 K, which is discussed elsewhere\cite{zhaozw}. From these MHLs one can calculate $j_c$ via $j_c = 20 \Delta M/Va(1-a/3b)$ based on the Bean critical state model, where V, a and b are the volume, width and length of the sample, respectively. The result of $j_c$ is shown in Fig.2. It is clear that the magnetic critical current density $j_c$ of our sample is rather high. For example, at T = 18 K and $\mu_0H$ = 1 T, we have $j_c = 2 \times 10^6 A / cm^2$, which is about one order of magnitude higher than that of high pressure synthesized bulk samples\cite{wencpl,wenphysicac,takano}. 

For investigating the flux dynamics the $j_c(H)$ curves have been measured with two different field sweeping rate 0.02 T / s and 0.01 T / s. It is interesting to note that although for bulk samples\cite{larbalestier,wenphysicac} there is a small tail on each MHL in high field region, here on the thin film sample there is no such tail. Accordingly the $j_c(H)$ curves show neither a small tail in high field region. This further corroborates our earlier suggestion that the small tail observed in bulk samples is due to some secondary effect, such as some local regions with very strong pinning or the surface pinning by the tiny grains. In the present thin film sample with much better uniformity this effect is certainly disappeared. From the contour of $j_c$ vs. H shown in Fig.2 one can see that all the curves have two different regions. In the low field region $j_c$ drops slowly with H. When the field is increased further and exceeded a threshold the $j_c$ decreases drastically showing the gradual setting in of the reversible flux motion. One can determine the phase transition line which separates the irreversible and the reversible flux motion by taking a criterion, here for example $j_c$ = 1000 $A / cm^2$. The same criterion was used by Bugoslavsky et al. for bulk samples\cite{bugoslavsky}. Since our measurement was done at a maximum field of 8 T, for low temperatures we use a reasonable extrapolating way, i.e., to follow the tendency of $j_c$ vs. H at a higher temperature, for example at 10 K or 14 K, down to the criterion $j_c$ = 1000 $A / cm^2$ to derive the irreversibility field. This method, although with some uncertainties for the values of $H_{irr}(T)$ at low temperatures, will lead us to derive an almost complete curve of $H_{irr}(T)$. The error bar of $H_{irr}(T)$ at 2 K is about $ \pm $ 0.5 T. The irreversibility lines by following this method is shown in Fig.3 together with that from a high pressure synthesized bulk sample. Similar to what found in bulk samples \cite{finnemore,bugoslavsky,larbalestier,wencpl,wenphysicac}, it is easy to see in the present thin film that the $H_{irr}(T)$ extrapolates to a rather low field, for example, $H_{irr}(0)\approx 9.2\pm$ 0.5 T, while the $H_{c2}(T)$ extrapolates to a much higher value ( $\approx$ 15 T )\cite{bud1} at zero K. There is a large separation between the two fields $H_{c2}(0)$ and $H_{irr}(0)$. This effect observed both in bulk and thin film samples may suggest that the relatively low $H_{irr}(T)$ in $MgB_2$ is not due to the easy flux motion through some weak pinning channels, rather it reflects probably a more intrinsic property, specialy in a rather clean system.

\section{Discussion}

\subsection {Large separation between $H_{irr}(0)$ and $H_{c2}(0)$ and possible evidence for the quantum vortex liquid in $MgB_2$}

If following the hypothesis of the vortex liquid above $H_{irr}(T)$, we would conclude that there is a large region for the existence of a quantum vortex liquid at zero K. This can be attributed to a quantum fluctuation effect of vortices in bulk $MgB_2$. Although the lowest temperature in our present experiment is 2 K, however, from the experimental data one cannot find any tendency for $H_{irr}(T)$ to turn upward to meet the $H_{c2}(0)$ at zero K. This may indicate that the vortex melting in our present film is due to the strong quantum fluctuation which smears the perfect vortex lattice leading to the vanishing of the shear module $C_{66}$ of the vortex matter ( probably within grains ). Dense disorders will strengthen the shear module and thus enhance the irreversibility line, that's why the $H_{irr}$ in present thin film is higher than that in bulks ( shown in Fig.3 ). The irradiation of protons by Bugoslavsky et al.\cite{perkins} did not suppress but strongly increase $j_c$ at a high field would suggest that the low value of $ H_{irr}(T) $ measured in unirradiated bulks and present thin film is not due to the weak links since otherwise the $j_c$ value would drop even faster with increasing the magnetic field after the irradiation. 

In order to investigate the flux dynamics in the vortex solid state below $H_{irr}$ we have carried out the dynamical magnetic relaxation measurement. Assuming an uniform current density over the cross-section of a superconducting ring, one can determine the superconducting current density $j_c$ from the magnetic moment M via

\begin{equation}
M = \frac{1}{3}\pi j_c d({R_o}^3-{R_i}^3) 
\end{equation}

where d is the film thickness, $R_o$ and $R_i$ are the outer and inner radius of the ring, respectively. It is important to note that the magnetic moment M in Eq.(1) is understood as being due to only the superconducting current excluding any additional contribution from an equilibrium magnetization. In other words, M is obtained by subtracting from the ZFC magnetic moment with the equilibrium magnetic moment as determined from the FC process, i.e., M = M ( measured ) - M ( equilibrium ). For a superconducting thin film the equilibrium magnetic moment is normally negligible since the volume of the superconducting material is very small. Based on different external conditions, there are two techniques to measure the magnetic relaxation, namely the so-called conventional and dynamical relaxation. The so-called conventional relaxation is to measure the time dependence of the superconducting current density $j_c$ at a certain temperature and field \cite{yeshurun,thompson2,kung}.  After a waiting time when the magnetic field is fixed ( or say field sweeping is stopped ), the first data point is taken. For a conventional relaxation measurement, total observation time should be very long. The second method is the so-called dynamical relaxation, i.e., to measure the MHL with different field sweeping rate. One can easily understand the difference between these two methods from the electromagnetic response of a superconducting ring. The electromotive force in a ring is

\begin{equation}
E \times 2\pi R=\pi R^2 \times \frac{d ( \mu_0 H)}{dt}-wdL \frac{dj_c}{dt} 
\end{equation}

where E is the electric field established within the ring, $R = ( R_o+ R_i )/2$, H is the external field, $ L= \mu_0R[ln(8R/w)-1/2]$ is the self-inductance of the ring \cite{brandt}, $w ( = R_o-R_i )$ is the width of the ring. For the relaxation process, since $dH/dt$ = 0, the electric field can be determined by 

\begin{equation}
E = - \frac { \mu_0wd}{2\pi}[ln(\frac{8R}{w})-\frac{1}{2}]\frac{dj_c}{dt}
\end{equation}

This method is inapplicable when the irreversible magnetic signal is comparable to the equilibrium magnetization. For $MgB_2$ with a very narrow MHL this method shows a clear drawback. However one can choose the so-called dynamical relaxation method\cite{jirsa,schnack}. This technique can be accomplished by using the sensitive VSM or a torque magnetometer. In a field sweeping process, if the field sweeping rate is high enough, the last term in eq.(2) can be neglected, therefore the electric field E can be determined by 

\begin{equation}
E = \frac{R}{2} \frac{d(\mu_0H)}{dt}
\end{equation}

and $j_c$ can be determined from the width of the micro-hysteresis loops around a certain field via $j_c = 20 \Delta M/Va(1-a/3b)$ based on the assumption that the current density $j_c$ is uniform throughout the cross section of the ring \cite{vanderBeek}. Since the vortices are forced to move by the external field sweeping, this process is thus called as the dynamical relaxation \cite{jirsa,schnack}. For a superconducting disc the magnetic moment is contributed mainly from the current circulating near the perimeter of the sample. For example, the 7/8 of the total magnetic moment is contributed by the current flowing from 1/2R to R of a superconducting disc, where R is the radius of the disc. Therefore from a rough estimation, it is safe to derive the E(j) or V(I) relation for a disc sample by using Eq.(4). As indicated in\cite{schnack} and \cite{jirsa},  the normalized relaxation rate can be determined via $Q = dlnj_s/dlnE$ in the dynamical relaxation process, and it should be identical to $ S = -dlnM/dlnt$ determined in the conventional relaxation process. Later on Wen et al.\cite{wenprb95}, Perkins and Caplin \cite{perkins2} and Ji et al. \cite{ji} have shown that the conventional relaxation, dynamical relaxation and the DC transport method should give the same information of flux motion although the voltage range is different. ( voltage in the dynamical relaxation is much higher than that in the conventional relaxation ). 

The raw data with two different field sweeping rate ( 0.02 T / s, 0.01 T / s) are shown in Fig.2. The Q values vs. field for different temperatures are determined and shown in Fig.4. It is clear that the relaxation rate increases monotonically with the external magnetic field and extrapolates to 100\% at about the melting point $H_{irr}$. At 2 K it is found from the Q(H) data that the melting field ( where Q = 1 ) is about $8.7\pm0.5$ T, being close to $H_{irr}( T = 2 K )\approx 9.2 T\pm 0.5$ determined from the $j_c(H)$ curve. It is known that the $H_{irr}(T)$ is rather stable in low temperature region, therefore we can anticipate a rather low value of $H_{irr}(0)$ which is below 10 T being much lower than $H_{c2}(0)$. As already pointed out in our earlier publication\cite{wencpl}, the large separation between $H_{irr}^{bulk}(0)$ and $H_{c2}(0)$ may manifest the existence of the quantum vortex liquid due to strong quantum fluctuation of vortices in the pure system of $MgB_2$.

Theoretically, quantum melting of the vortex solid has been proposed by some authors\cite{blatter,kramer,ikeda} and preliminarily verified by experiments.\cite{sasaki,okuma}. Solid evidence is, however, still lacking mainly because either the values of $H_{irr}(0)$ and $H_{0}(T)$ are too high to be accessible, such as in the classical Chevrel phase PbMoS system\cite{rossel}, or the separation between them is too small \cite{sasaki,okuma} leading to a difficulty in drawing any unambiguous conclusions. Here we try to have an rough consideration on the quantum melting field $H_m$ proposed by Blatter et al.\cite{blatter} for 2D system, 
 
\begin{equation}
H_m(0)/H_{c2}(0) = 1-1.2exp(-\pi^3C_L^2R_Q/4R_{2D}) 
\end{equation}

where $C_L$ is the Lindermann number, $R_Q = \hbar/ e^2 \approx 4.1 k\Omega $, $R_{2D}$ is the sheet resistance. Since the new $MgB_2$ sample has a much higher charge density and thus a much lower sheet resistivity, according to above relation, $H_m$ should be more close to $H_{c2}(0)$ comparing to $high-T_c$ superconductors ( HTS ). This is in contrast to the experimental observations which may be explained as that the $MgB_2$ is not a quasi-2D system. Another approach was proposed by Rozhkov and Stroud\cite{rozhkov}, 

\begin{equation}
H_m(0)/H_{c2}(0) = B_0/(B_0+H_{c2}(0)) 
\end{equation}

with $B_0=\beta m_pC^2s\Phi_0/4\pi\lambda (0)^2q^2$, where s is the spacing between layers, $m_p$ is the pair mass, q the pair charge ( $\approx 2e$ ), C the light velocity, $\lambda(0)$ the penetration depth at zero K, $\beta \approx$ 0.1. If comparing again the present new superconductor $MgB_2$ with HTS, $\lambda(0)$, q and $\beta$ are more or less in the same scale, the difference comes from $m_p$ and s. Therefore a preliminary conclusion would be that in $MgB_2$ either the pair mass $m_p$ or the layer spacing s is much smaller than that of HTS.

\subsection{Residual relaxation rate at zero K and weak temperature dependence of the relaxation rate}

A strong quantum fluctuation normally favors a strong quantum tunneling creep. In order to see that, we plot in Fig.5 the temperature dependence of the relaxation rate Q. It is clear that there is a clear residual relaxation rate for all fields and the relaxation rate in wide temperature region keeps rather stable against the thermal activation and fluctuation. According to the thermally activated flux motion model

\begin{equation}
E=v_0Bexp(- \frac{U(j_c,T)}{k_BT})
\end{equation}

where E is the electric field due to TAFM over the activation energy $U(j_c,T)$, $v_0$ is the average velocity of the flux motion, B is the magnetic induction. The relaxation rate is

\begin{equation}
Q = \frac{dlnj_c}{dlnE}= -k_BT(\frac{j_cdU(j_c,T,B)}{dj_c})
\end{equation}

For any kind of $U(j_c)$ relation, a finite slope of $dU/dj_c$ is expected. Therefore a much stronger temperature dependence of Q should be expected for thermally activated flux motion. This is in contrast to the experimental data. However, when the melting point $H_{irr}(T)$ is approached the relaxation rate will quickly jump to 100\%. This may indicate that the thermal fluctuation is not the dominant process for the flux depinning in the new superconductor $MgB_2$. It shows high possibility for the vortex quantum melting even at a finite temperature. Worthy to note is that the quantum tunneling rate is extremely low at an intermediate field, such as Q = 0.3\% at 2 K and 2 T, but is rather high at a high field, for example Q = 20\% at 2 K and 7 T. This may imply that the field will greatly enhance the vortex quantum fluctuation and tunneling.

The small relaxation rate at a relatively low field has also been measured by Thompson et al.\cite{thompson} who regarded it as a highly stable superconducting current density in $MgB_2$. Actually the relaxation rate can be rather high when the magnetic field is increased to a higher value. The extremely small relaxation rate and weak temperature dependence at a finite temperature at a low field is probably induced by a strong pinning barrier relative to the thermal energy, i.e., $k_BT << U_c$, where $U_c$ is the intrinsic pinning energy. Recently it was concluded\cite{zhaozw2,jinhao} that the $U_c$ is in the scale of 1000 K being much higher than the thermal energy $k_BT$. Therefore for the new superconductor $MgB_2$ the pinning well is too deep leading to a trivial influence of the thermal activation and fluctuation. It thus naturally suggests that the quantum fluctuation and tunneling plays an more important role. Therefore, together with the fact discussed in last subsection, it is tempting to suggest that at a finite temperature the melting between a vortex solid and a liquid is due to quantum fluctuation instead of the thermal fluctuation. 

\section{Conclusion}
In conclusion, in a thin film sample of $MgB_2$, the flux dynamics and the irreversibility field are investigated. Just like in the bulk samples, is is found that the irreversibility field is rather low comparing to the upper critical field in low temperature region showing the possible existence of the quantum vortex liquid due to strong quantum fluctuation. The weak temperature dependence but strong field dependence of the relaxation rate may further suggest that the vortex melting at a finite temperature is also induced by the strong quantum fluctuation.  The reason for such a strong quantum effect is still unknown, but it may be related to the superconducting mechanism of $MgB_2$, such as the relatively low upper critical field.   

\acknowledgements
This work is supported by the National Science Foundation of China (NSFC 19825111) and the Ministry of Science and Technology of China ( project: NKBRSF-G1999064602 ). HHW gratefully acknowledges Prof. B. Ivlev and Dr. A. F. Th. Hoekstra for fruitful discussions, and continuing financial support from the Alexander von Humboldt foundation, Germany.

\begin{figure}
\caption{Magnetization hysteresis loops measured at 2, 4, 6, 8, 10, 14, 18, 22, 26, 30, 32, 35, 37, K and 38 K ( from outer to inner ). All curves here show a symmetric behavior indicating the importance of bulk current instead of surface shielding current. The MHLs measured at low temperatures ( e.g., 2 K to 10 K ) are too close to be distinguishable. Dense and small flux jumps have been observed below 10 K near the central peak.
}
\end{figure}

\begin{figure}
\caption{Critical current density $j_c$ calculated based on the Bean critical state model. At each temperature the data has been measured with two field sweeping rate: 0.02 T / s and 0.01 T / s. The faster sweeping rate corresponds to a higher dissipation and thus higher current density. From these data one can calculate the dynamical magnetic relaxation rate Q. The $j_c(H)$ curves measured at low temperatures are very close to each other showing a rather stable value of $H_{irr}$ when T approaches zero K. We use a criterion of $j_c = 1000 A/cm^2$ to determine the irreversibility line.
}
\end{figure}

\begin{figure}
\caption{
H-T phase diagram for the new superconductor $MgB_2$. The circles represent the bulk irreversibility lines $ H_{irr}$ of bulk samples: open and filled circles represent two measurement  by VSM on two bulk samples. The open diamond symbols represent the irreversibility line of the thin film. The squares represent the upper critical field $H_{c2}(T)$: filled squares are from resistive measurement; open squares are from the M(T) measurement by SQUID. All lines are guide to the eye.}
\end{figure}

\begin{figure}
\caption{
Field dependence of the relaxation rate at temperatures of 2, 4, 6, 8, 10, 14, 18, 20, 22, 24, 26, 28, 30, 32, 35 K. The dashed line is a guide to the eye for 2 K. It is clear that Q will rise to 100\% at about 8.7 T at 2 K. Since $H_{irr}$ is rather stable at low temperatures, it is safe to anticipate that $H_{irr}(0) < 10 T$ being much smaller than $H_{c2}(0)\approx $ 15 T.}
\end{figure}

\begin{figure}
\caption{
Temperature dependence of the relaxation rate at fields of 1 to 8 T with increments of 1 T. A clear residual relaxation rate is observed at all magnetic fields. The relaxation is weakly dependent on temperature until the irreversibility temperature is reached. These are difficult to be understood in a frame of thermal acitivation flux motion and thermal depping. }
\end{figure}


\begin{references}

\bibitem[*]{response} E-mail address: hhwen@aphy.iphy.ac.cn.
\bibitem{akimitsu} J. Nagamatsu, N. Nakagawa, T. Maranaka, Y. Zenitani and J. Akimitsu, Nature 410, 63(2001).
\bibitem{bugoslavsky} Y. Bugoslavsky, G. K. Perkins, X. Qi, L. F. Cohen, and A. D. Caplin, Nature 410, 563(2001).
\bibitem{finnemore} D. K. Finnemore, J. E. Ostenson, S. L. Bud'ko, G. Lapertot, and P. C. Canfield, Cond-mat / 0102114.
\bibitem{larbalestier} D. C. Larbalestier, et al., Nature 410, 186(2001).
\bibitem{thompson} J. R. Thompson, M. Paranthaman, D. K. Christen, K. D. Sorge, H. J. Kim and J. G. Ossandon, Cond-mat / 0103514 ( 2001 ).
\bibitem{wencpl} H. H. Wen, S. L. Li, Z. W. Zhao, Z. A. Ren, G. C. Che and Z. X. Zhao, Chin. Phys. Lett. 86, 816 ( 2001 ).
\bibitem{wenphysicac} H. H. Wen, S. L. Li, Z. W. Zhao, Z. A. Ren, G. C. Che and Z. X. Zhao, submitted to Physica C.
\bibitem{qinmj} M. J. Qin, X. L. Wang, H. K. Liu, S. X. Dou, Cond-mat / 0104112 ( 2001 ).
\bibitem{dhalle} M. Dhalle, P. Toulemonde, C. Beneduce, N. Musolino, M. Decroux, Cond-mat / 0104395 ( 2001 ).
\bibitem{kang} W. N. Kang {\it et al.}, Science in press, cond-mat/0103179 (2001).
\bibitem{zhaozw}Z. W. Zhao, H. H. Wen, S. L. Li, Y. M. Ni,H. P. Yang, W. N. Kang, H. J. Kim, E. M. Choi and S. I. Lee, Cond-mat / 0104249 ( 2001 ). 
\bibitem{takano} Y. Takano, H. Takeya, H. Fujii, H. Kumakura, T. Hatano, K. Togano, H. Kito, and H. Ihara, Cond-mat / 0102167.
\bibitem{bud1} S. L. Bud'ko, C. Petrovic, G. Lapertot, C. E. Cunningham, and P. C. Canfield, Cond-mat / 0102413.
\bibitem{perkins} Y. Bugoslavsky {\it et al.}, cond-mat/0104156(2001).
\bibitem{yeshurun} Y. Yeshurun, A. P. Malozemoff, A. Shaulov, Rev. Mod. Phys. {\bf 68}, 911 (1996). 
\bibitem{thompson2} J. R. Thompson, Yang Ren Sun, L. Civale, A. P. Malozemoff, M. W. McElfresh, A. D. Marwick and F. Holtzberg, Phys.Rev.B {\bf 47}, 14440(1993). J. R. Thompson, Yang Ren Sun, D. K. Christen, Phys. Rev. B {\bf 49}, 13287(1994).
\bibitem{kung} P. J. Kung, M. P. Maley, M. E. McHenry, and J. O. Willis, M. Murakami, and S.Tanaka, Phys. Rev. B {\bf 48}, 13922(1993).
\bibitem{brandt} J. Gilchrist and E. H. Brandt, Phys. Rev. {\bf B 54}, 3530(1996).
\bibitem{jirsa} M. Jirsa, L. Pust, H. G. Schnack, R. Griessen, Physica {\bf C 207}, 85(1993).
\bibitem{schnack} H. G. Schnack, R. Griessen, J. G. Lensink, C. J. van der Beek, and P. H. Kes, Physica {\bf C 197}, 337(1992).
\bibitem{vanderBeek} C. J. van der Beek, G. J. Nieuwenhuys, P. H. Kes, H. G. Schnack, and R. Griessen, Physica {\bf C 197}, 320(1992).
\bibitem{wenprb95} H. H. Wen, H. G. Schnack, R. Griessen, B. Dam, and J. Rector, Physica C {\bf241}, 353(1995).
\bibitem{perkins2} G. K. Perkins, A. D. Caplin, Phys. Rev. {\bf B 54}, 1255(1996).
\bibitem{ji} H. L. Ji, Z. X. Shi, Z. Y. Zeng, X. Jin, X. X. Yao, Physica {\bf C217}, 127(1993).
\bibitem{blatter} G. Blatter, and B. Ivlev, Phys. Rev. Lett. 70, 2621 ( 1993 ). G. Blatter, and B. I. Ivlev, Phys. Rev. B 50, 10272 ( 1994 ). G. Blatter, et al. Phys. Rev. B 50, 13013 ( 1994 ).
\bibitem{kramer} A. Kramer, and S. Doniach, Phys. Rev. Lett. 81, 3523 ( 1998 ).
\bibitem{ikeda} R. Ikeda, Int. J. Mod. Phys. B 10, 601 ( 1996 ).
\bibitem{sasaki} T. Sasaki, W. Biberacher, K. Neumaier, W. Hehn, A. Andres, Phys. Rev. B 57, 10889 ( 1998 ). 
\bibitem{okuma} S. Okuma, Y. Imamoto, and M. Morita, Phys. Rev. Lett.86, 3136(2001).
\bibitem{rossel} C. Rossel, E. Sandvold, M. Sergent, R. Chevrel, and M. Potel, Physica C 165, 233(1990). C. Rossel, O. Pena, H. Schmitt, and M. Sergent, Physica C181, 363 ( 1991 ).
\bibitem{rozhkov} A. Rozhkov, and D. Stroud, Phys. Rev. B 54, R12697 ( 1996 ).
\bibitem{zhaozw2} Z. W. Zhao, H. H. Wen, S. L. Li, Y. M. Ni, Z. A. Ren, G. C. Che, H. P. Yang, Z. Y. Liu and Z. X. Zhao, Chinese Physics 10, 340(2001).
\bibitem{jinhao} H. Jin, H. H. Wen, S. L. Li, Z. W. Zhao, Y. M. Ni, Z. A. Ren, G. C. Che, H. P. Yang, Z. Y. Liu and Z. X. Zhao, Chin. Phys. Lett. 18, 823 ( 2001 ). 
\end{references}
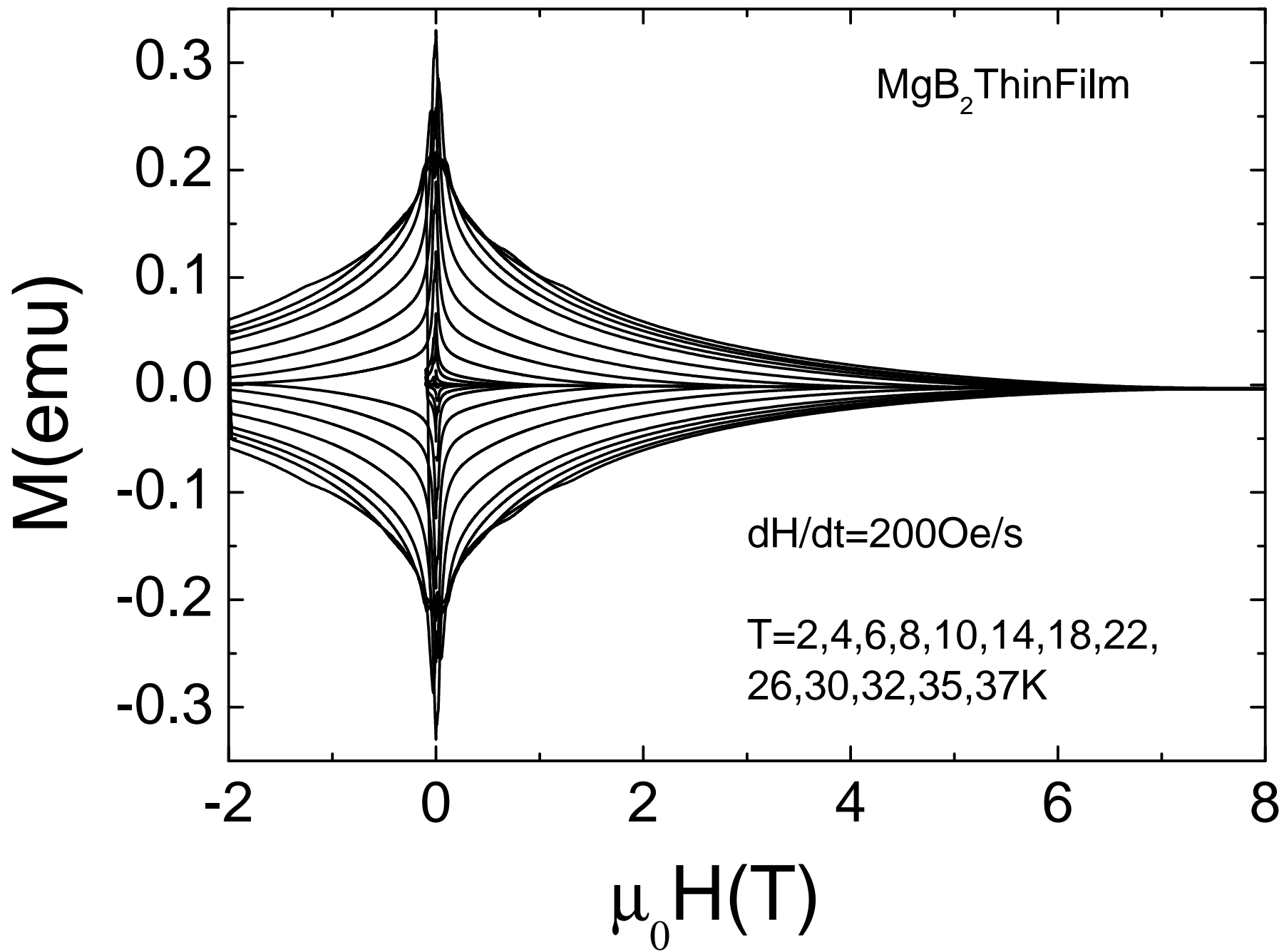
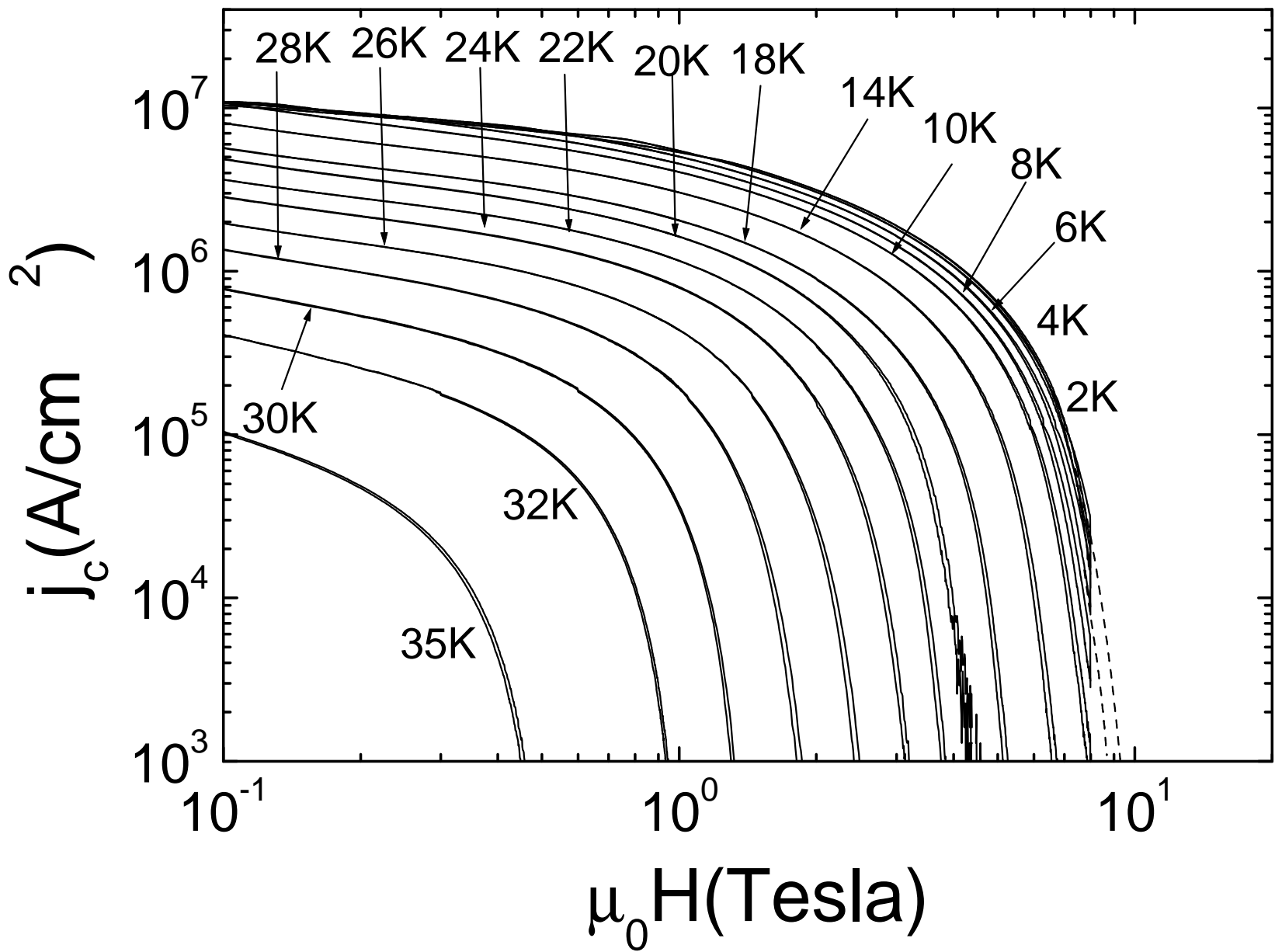
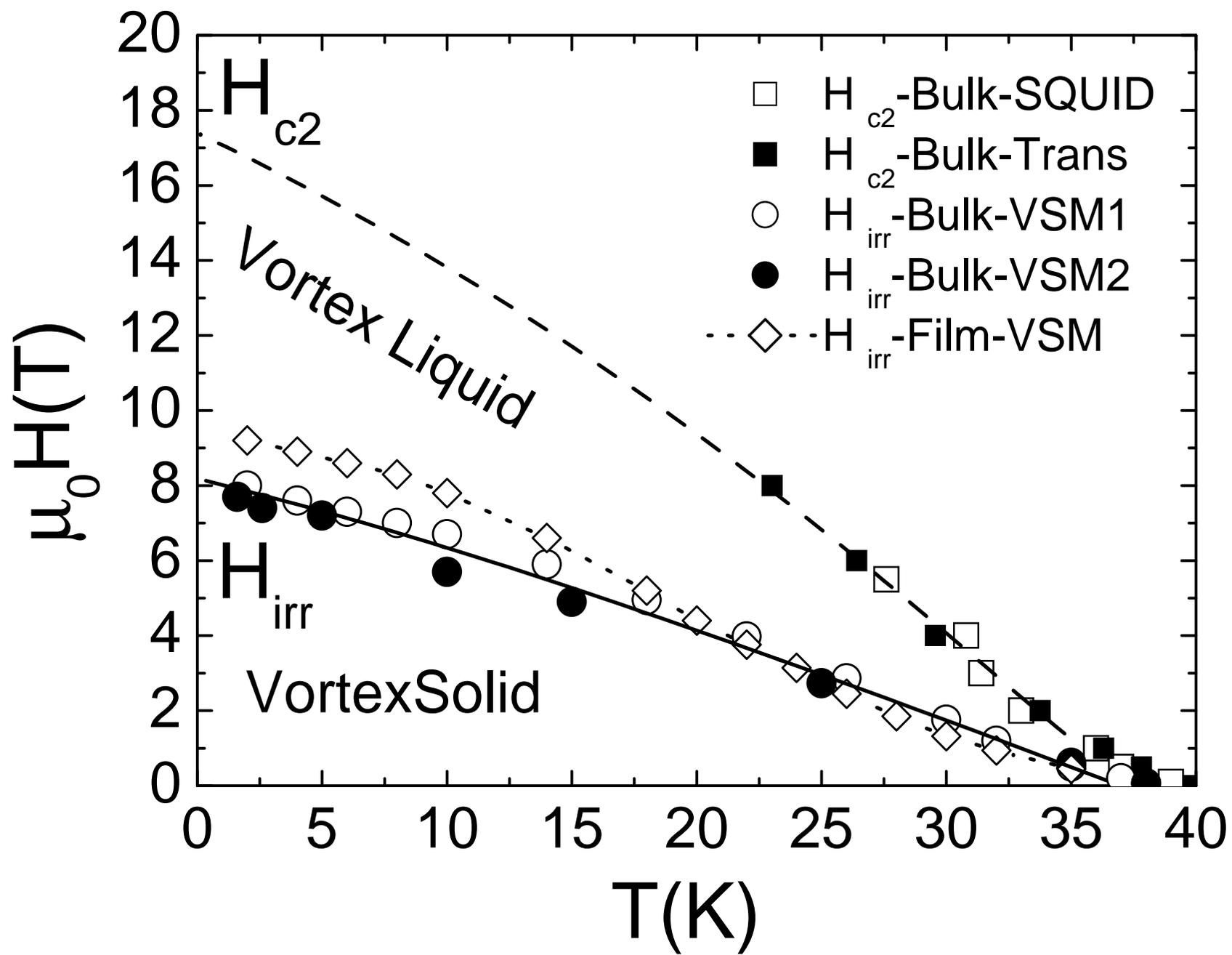
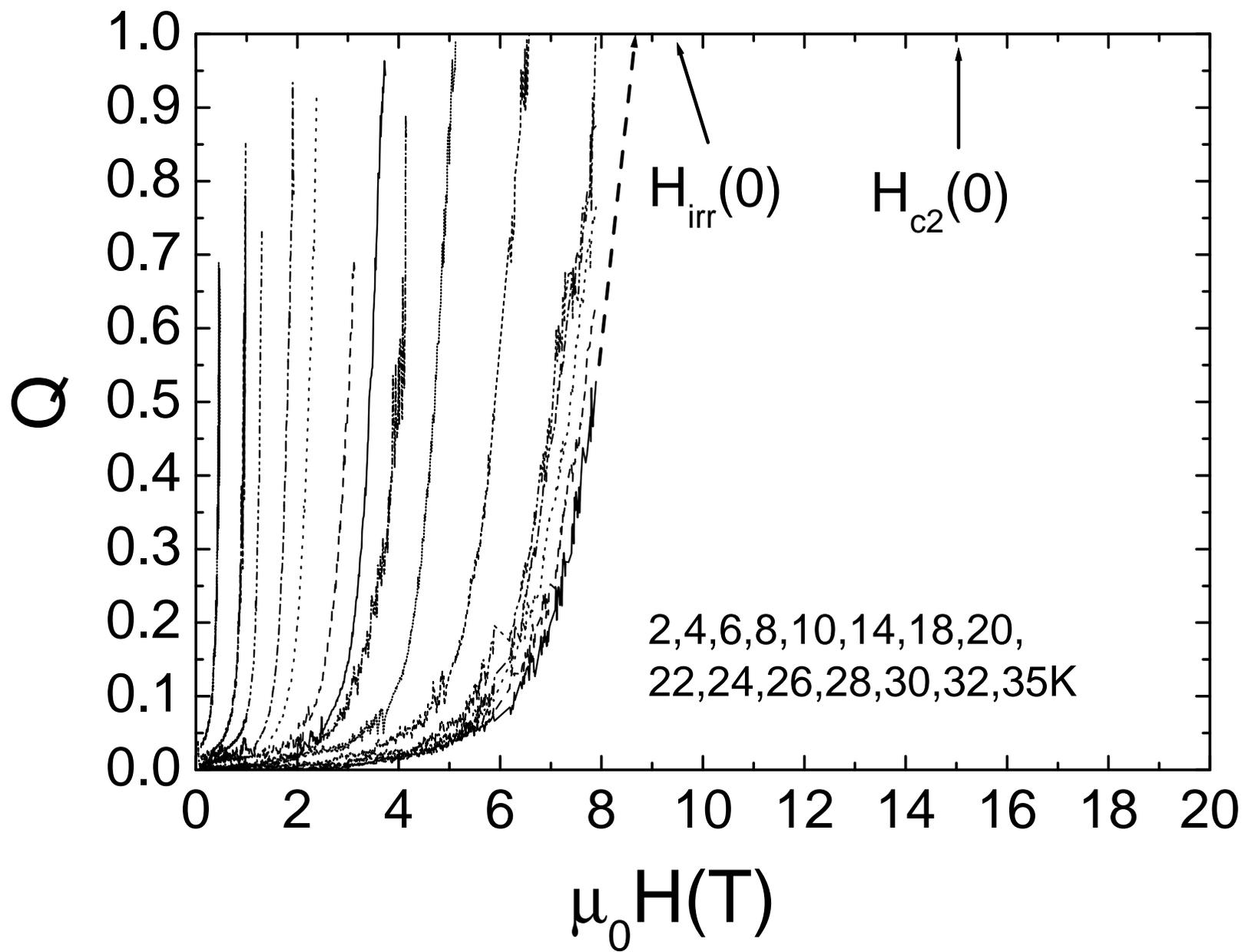
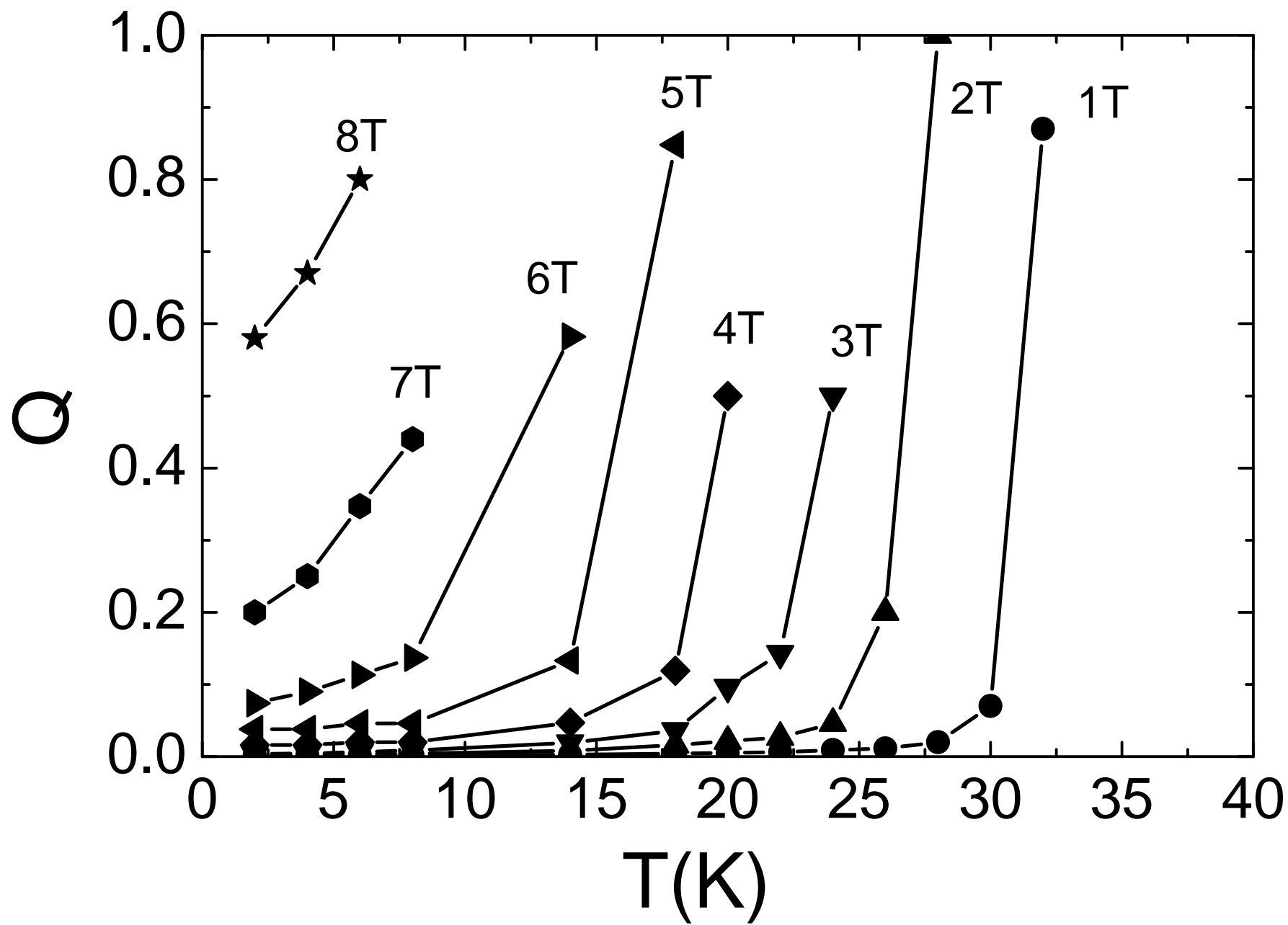
\end{document}